\documentclass[11pt,english]{article}
\usepackage{times}
\usepackage[T1]{fontenc}
\usepackage{a4wide}
\usepackage{amsmath}
\usepackage{graphicx}
\usepackage{setspace}
\usepackage{amssymb}
\usepackage[numbers]{natbib}

\makeatletter

\newcommand{\noun}[1]{\textsc{#1}}

\usepackage{babel}
\makeatother
\begin{document}

\title{Bayesian analysis of signal deconvolution \\
using measured instrument response functions}

\author{Pascal \noun{Pernot}\\Laboratoire de Chimie Physique, (UMR 8000,
associated to CNRS) \\Bât. 349, Université Paris-Sud, 91405 Orsay
Cedex, France\\email: pascal.pernot@lcp.u-psud.fr}

\maketitle
\begin{abstract}
Using measured instrumental response functions for data deconvolution
is a known source of uncertainty. This problem is revisited here with
Bayesian data analysis an Monte Carlo simulations. Noise correlation
induced by the convolution operator is identified as a major source
of uncertainty which has been neglected in previous treatments of
this problem. Application to a luminescence lifetime measurement setup
shows that existing approximate treatments are markedly defficient
and that the correlation length of the noise is directly related to
the lifetime to be estimated. Simple counteractive treatments are
proposed to increase the accuracy of this procedure.
\end{abstract}
\newpage

\section{Introduction}

The deconvolution problem is a classical inverse problem, and has
received a lot of attention in many scientific and engineering fields.
The instrument response function (IRF), also called blurring function,
is generally assumed to be accurately determined. It is however not
uncommon that the IRF is measured with the same accuracy as the signal
to be treated, due to instrumental or experimental design constraints.
The impact of an uncertain IRF on the accuracy of the deconvolved
signal has to be considered with care. This uncertainty propagation
issue has been addressed in the past by Dose \emph{et al.} \citep{Dose98}.
We show here analytically and numerically that their approximate solution
does not encompass important effects of noise correlation due to convolution. 

In this paper, we use Bayesian data analysis to derive an exact expression
of the likelihood function in the case of gaussian additive noise.
This solution is applied to the classical problem of lifetime estimation
from luminescence data.

\section{Theory}

\noindent The observed signal vector $\mathbf{s}$ (length $n$) is
generally expressed as a linear reconvolution model \begin{equation}
\mathbf{s}=\mathbf{Hm}+\mathbf{e}_{s},\label{eq1}\end{equation}
where $\mathbf{m}$ is a vector of values of the model function at
the measurement points, $\mathbf{H}$ is a $n\times n$ zero-padded
lower triangular Toeplitz matrix built from the IRF $\mathbf{h}$
of length $n_{h}$ \begin{equation}
\mathbf{H}=\left(\begin{array}{ccccc}
h_{1} & 0 & \cdots & 0 & 0\\
h_{2} & h_{1} & \cdots & 0 & 0\\
\vdots & \vdots & \ddots & 0 & 0\\
h_{n_{h}} & h_{n_{h}-1} &  & h_{1} & 0\\
0 & h_{n_{h}} & \cdots & h_{2} & h_{1}\end{array}\right)\label{mat}\end{equation}
and $\mathbf{e}_{s}$ is an additive noise with multinormal statistics
and covariance matrix $\mathbf{R}_{s}$: \begin{equation}
\mathbf{e}_{s}\sim\mathcal{N}_{n}(0,\mathbf{R}_{s}).\label{eq:noise}\end{equation}
 Note that we use bold lowercase symbols for vectors ($\mathbf{s}$)
and bold capitals for matrices ($\mathbf{H}$).

\noindent Using the symmetry property of convolution, Eq. \ref{eq1}
can also be written\begin{equation}
\mathbf{s}=\mathbf{Mh}+\mathbf{e}_{s}\label{eq2}\end{equation}
where $\mathbf{M}$ is a $n\times n_{h}$ lower triangular Toeplitz
matrix built from the model vector $\mathbf{m}$ as shown above (Eq.
\ref{mat}). 

\noindent As the exact IRF is generally not known, a measured IRF
is used instead to solve Eq.\ref{eq1} or Eq.\ref{eq2}. A multinormal
additive noise model is also used for the IRF\begin{equation}
\mathbf{h}=\hat{\mathbf{h}}+\mathbf{e}_{h},\end{equation}
where $\hat{\mathbf{h}}$ is the unknown exact IRF and $\mathbf{e}_{h}\sim\mathcal{N}_{n_{h}}(0,\mathbf{R}_{h}).$

The problem is to reconstruct the model vector $\mathbf{m}$, knowing
$\mathbf{s}$, $\mathbf{R}_{s}$, $\mathbf{h}$ and $\mathbf{R}_{h}$,
and to evaluate the impact of the measurement uncertainties of $\mathbf{h}$
on $\mathbf{m}$. This is an inverse problem doubled with an uncertainty
propagation problem. Bayesian data analysis is very well suited to
handle this kind of problem \citep*{Gelman95,Sivia96,Press89,Hanson99a}.

\subsection{\noindent Bayesian data analysis}

The posterior probability density function (pdf) for $\mathbf{m}$
is obtained by Bayes's formula\begin{equation}
p(\mathbf{m}|\mathbf{s},\mathbf{R}_{s},\mathbf{h},\mathbf{R}_{h})=\frac{p(\mathbf{m})}{p(\mathbf{s})}p(\mathbf{s}|\mathbf{m},\mathbf{R}_{s},\mathbf{h},\mathbf{R}_{h}),\end{equation}
where $p(\mathbf{m})$ and $p(\mathbf{s})$ are the prior pdf's for
$\mathbf{m}$ and $\mathbf{s}$, and where $p(\mathbf{s}|\mathbf{m},\mathbf{R}_{s},\mathbf{h},\mathbf{R}_{h})$
is the likelihood function. Given our model, we do not know explicitely
this latter function. Instead, we know the explicit expression for
the likelihood when the exact IRF $\hat{\mathbf{h}}$ is considered
(cf. Eq. \ref{eq:noise})\begin{equation}
p(\mathbf{s}|\mathbf{m},\mathbf{R}_{s},\hat{\mathbf{h}})\sim\mathcal{N}_{n}(\mathbf{M}\hat{\mathbf{h}},\mathbf{R}_{s}).\end{equation}
Applying the marginalization rule and knowing the expression of the
pdf for $\hat{\mathbf{h}}$\begin{equation}
p(\hat{\mathbf{h}}|\mathbf{h},\mathbf{R}_{h})\sim\mathcal{N}_{n}(\mathbf{h},\mathbf{R}_{h}),\end{equation}
we can write\begin{equation}
p(\mathbf{m}|\mathbf{s},\mathbf{R}_{s},\mathbf{h},\mathbf{R}_{h})=\frac{p(\mathbf{m})}{p(\mathbf{s})}\int d\hat{\mathbf{h}}\: p(\mathbf{s}|\mathbf{m},\mathbf{R}_{s},\hat{\mathbf{h}})p(\hat{\mathbf{h}}|\mathbf{h},\mathbf{R}_{h}).\end{equation}
Considering that in our model $p(\mathbf{s})$ is a normalization
constant, and expliciting the pdf's, one gets\begin{equation}
p(\mathbf{m}|\mathbf{s},\mathbf{R}_{s},\mathbf{h},\mathbf{R}_{h})\propto p(\mathbf{m})\int d\hat{\mathbf{h}}\:\exp\left(-\frac{1}{2}J\right),\end{equation}
where\begin{eqnarray}
J & = & (\mathbf{s}-\mathbf{M}\hat{\mathbf{h}})^{T}\mathbf{R}_{s}^{-1}(\mathbf{s}-\mathbf{M}\hat{\mathbf{h}})\nonumber \\
 & + & (\mathbf{h}-\hat{\mathbf{h}})^{T}\mathbf{R}_{h}^{-1}(\mathbf{h}-\hat{\mathbf{h}}).\end{eqnarray}
This quantity is rearranged in order to enable analytical integration\begin{eqnarray}
J & = & (\hat{\mathbf{h}}-\mathbf{h}_{0})^{T}\mathbf{P}^{-1}(\hat{\mathbf{h}}-\mathbf{h}_{0})-\mathbf{h}_{0}^{T}\mathbf{P}^{-1}\mathbf{h}_{0}\nonumber \\
 & + & \mathbf{s}^{T}\mathbf{R}_{s}^{-1}\mathbf{s}+\mathbf{h}^{T}\mathbf{R}_{h}^{-1}\mathbf{h},\end{eqnarray}
where\begin{equation}
\left\{ \begin{array}{rcl}
\mathbf{P} & = & \left(\mathbf{M}^{T}\mathbf{R}_{s}^{-1}\mathbf{M}+\mathbf{R}_{h}^{-1}\right)^{-1}\\
\mathbf{h}_{0} & = & \mathbf{P}(\mathbf{M}^{T}\mathbf{R}_{s}^{-1}\mathbf{s}+\mathbf{R}_{h}^{-1}\mathbf{h})\\
 & = & \mathbf{h}+\mathbf{PM}^{T}\mathbf{R}_{s}^{-1}(\mathbf{s}-\mathbf{Mh})\end{array}\right.\end{equation}
Integration over $\hat{\mathbf{h}}$ finally leads to\begin{equation}
p(\mathbf{m}|\mathbf{s},\mathbf{R}_{s},\mathbf{h},\mathbf{R}_{h})\propto\frac{p(\mathbf{m})}{|\mathbf{P}|^{1/2}}\exp\left(-\frac{1}{2}(\mathbf{s}-\mathbf{Mh})^{T}\mathbf{K}(\mathbf{s}-\mathbf{Mh})\right),\end{equation}
where\begin{equation}
\mathbf{K}=\mathbf{R}_{s}^{-1}-\mathbf{R}_{s}^{-1}\mathbf{MPM}^{T}\mathbf{R}_{s}^{-1}.\label{result}\end{equation}
This expression for the posterior pdf calls for a few comments : 

\begin{itemize}
\item \noindent convolution of the model vector with a noisy IRF leads to
a ''noisy model'' $\tilde{\mathbf{m}}=\mathbf{Mh}$, affected by
correlated noise with covariance matrix $\mathbf{K}^{-1}$, the structure
of which depends explicitely on the model vector itself (heteroscedastic
correlated noise). This is in contrast with the result of Dose \emph{et
al. \citep{Dose98}}, who obtain an expression for an effective variance,
and do not consider the covariance part. 
\item \noindent the mode of the posterior pdf depends on the actual value
of the measured IRF $\mathbf{h}$. A bias in the optimal values for
the model vector is thus to be expected, as a different realization
of the IRF would lead to a different solution. In any case, a consistent
uncertainty analysis should ensure that the exact value lies within
confidence intervals. 
\end{itemize}

\section{\noindent Application}

An application of interest is for instance the lifetime estimation
of unstable chemical species from their luminescence decays.

\subsection{Model}

A mono-exponential decay signal with lifetime $\tau$ is generated
over a regular time grid ($n_{h}=n=100$). The model is $m_{i}=\textrm{exp}(-t_{i}/\tau)$.
The IRF is a gaussian function centered at $t_{0}$, and of FWHM $w_{h}$
\begin{equation}
h_{i}=\textrm{exp}\left(-4\ln(2)(t_{i}-t_{0})^{2}/w_{h}^{2}\right).\end{equation}
 In order to keep a single parameters, the model after convolution
is rescaled to the maximal value of the signal, and we can set $p(\tau|\mathbf{s},\mathbf{R}_{s},\mathbf{h},\mathbf{R}_{h})\equiv p(\mathbf{m}|\mathbf{s},\mathbf{R}_{s},\mathbf{h},\mathbf{R}_{h})$.
Homoscedastic noise is considered for both signal and IRF, i.e. $\mathbf{R}_{s}=\sigma_{s}^{2}*\mathbf{I}_{n}$
, $\mathbf{R}_{h}=\sigma_{h}^{2}*\mathbf{I}_{n_{h}}$. Finally, a
uniform prior distribution for $\tau$ is used ($p(\tau)=cte$).

\begin{figure}[ht]
\begin{center}\includegraphics[%
  clip,
  scale=0.4]{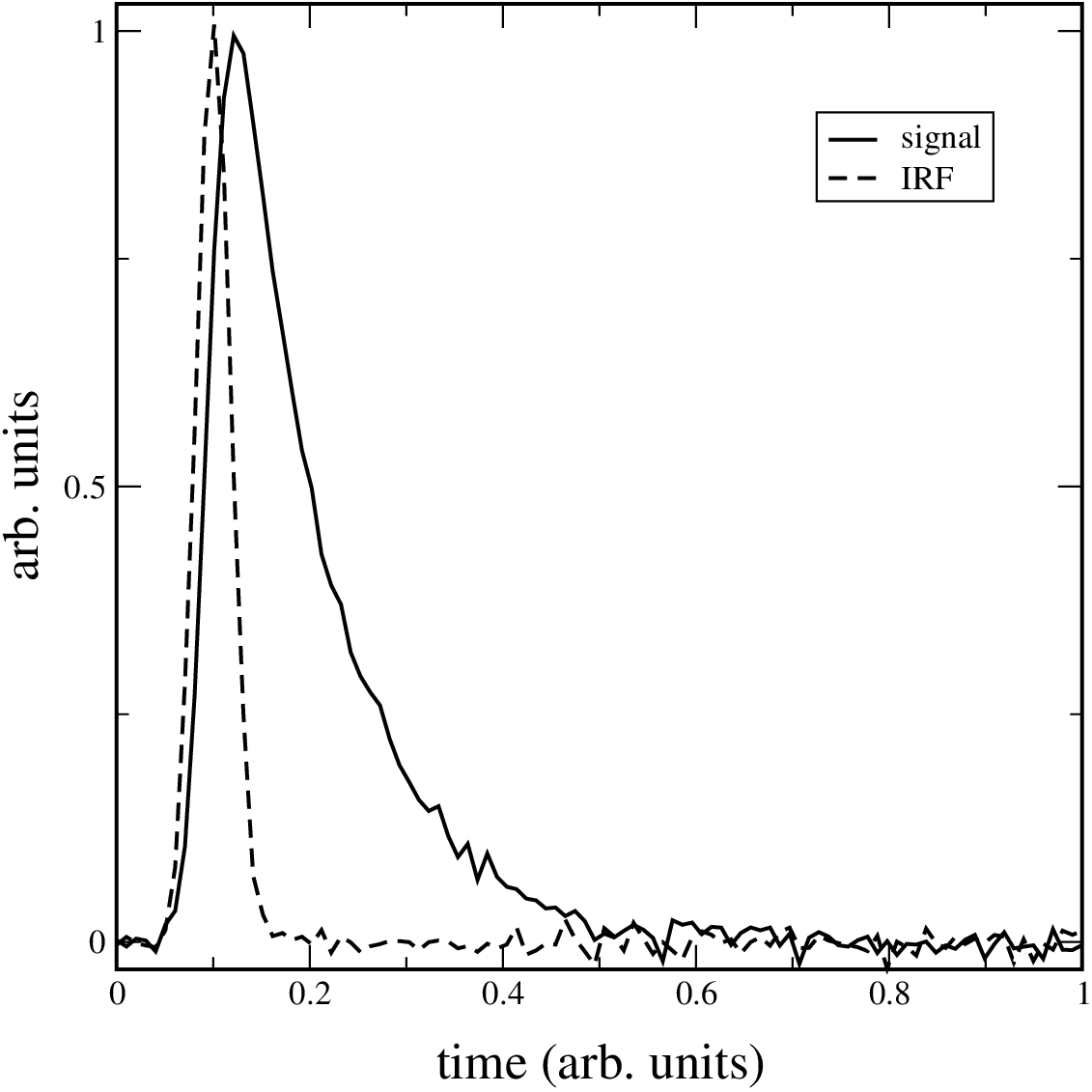}\end{center}

\caption{\label{fig_decay}Typical synthetic signal and IRF used for lifetime
estimation ($\sigma_{s}=\sigma_{h}=0.01$, $\tau=0.1$ $t_{0}=0.1$and
$w_{h}=0.03$).}
\end{figure}

\subsection{Comparison of models of the posterior pdf}

The exact expression for the posterior pdf (eq.\ref{result}) is compared
to approximate expressions :

\begin{itemize}
\item no correction for the noisy IRF ($\sigma_{h}=0$, in our model), which
is the most commonly used method;
\item the diagonal approximation of eq.\ref{result}, which implements some
level of variance correction, but fails to encompass the correlation
in the model's noise;
\item the ''effective variance'' method \citep{Dose98,Toussaint99}.
\end{itemize}
\begin{figure}[ht]
\begin{center}\includegraphics[%
  clip,
  scale=0.4]{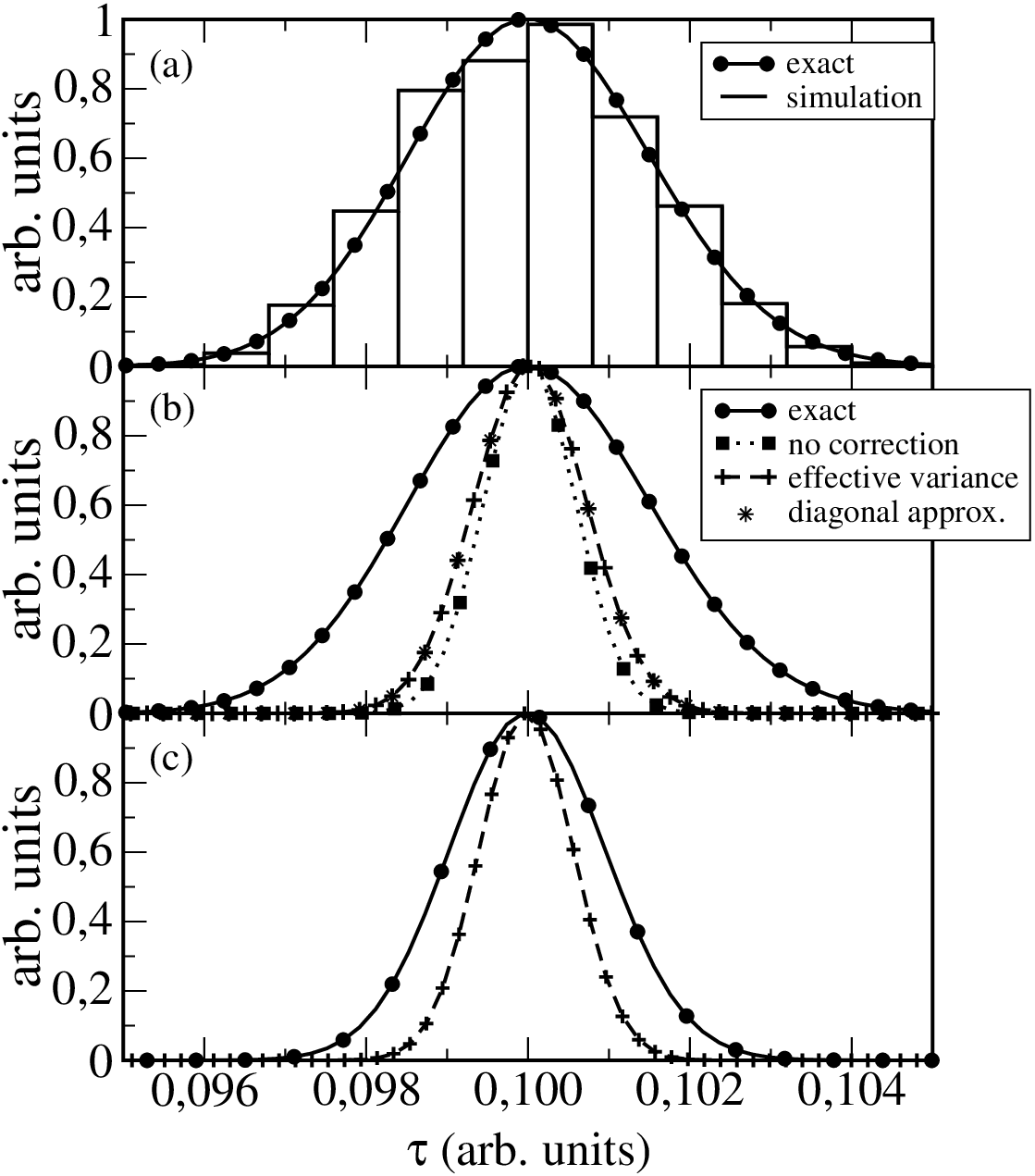}\end{center}

\caption{\label{fig_dist1}Posterior density functions for the lifetime estimated
from a noisy decay and for different treatments of the IRF's uncertainty
($\sigma_{h}=\sigma_{s}=0.01$). The functions have been shifted and
renormalized to facilitate direct comparison. (a) comparison of simulation
results (1000 runs) with the full bayesian solution proposed in the
present work; (b) comparison of the full treatment with various approximations
(see text), full support of the IRF; (c) support of the IRF limited
to $t\leq0.2$ (all approximate methods are undiscernable).}
\end{figure}

\textbf{Variance.} Fig. \ref{fig_dist1} represents the posterior
pdf $p(\tau|\mathbf{s},\sigma_{s},\mathbf{h},\sigma_{h})$ computed
by Monte Carlo simulation, and by the different methods in the case
of a same measurement accuracy for the signal and the IRF ($\sigma_{h}=\sigma_{s}=0.01$).
The Monte Carlo method consists in repeated analysis of randomly noised
signal and IRF to build histograms of the maximum a posteriori (MAP)
lifetime values (modes of the posterior pdf). All curves have been
shifted to a common mode, in order to facilitate comparison. The exact
expression is fully coherent with the histogram resulting of the simulation,
i.e. it takes properly the variance of the signal and the variance
of the IRF into account. It can be seen on this figure that the approximate
methods all perform quite similarly and fail to recover the full variance
of the lifetime. The ''effective variance'' method is seen to be
numerically equivalent to the diagonal approximation of our method,
and it performs only slightly better than the totally uncorrected
method. Correlation in the noise of the convolved model can thus have
a major impact on uncertainty quantification.

\begin{figure}[ht]
\begin{center}\includegraphics[%
  clip,
  scale=0.4]{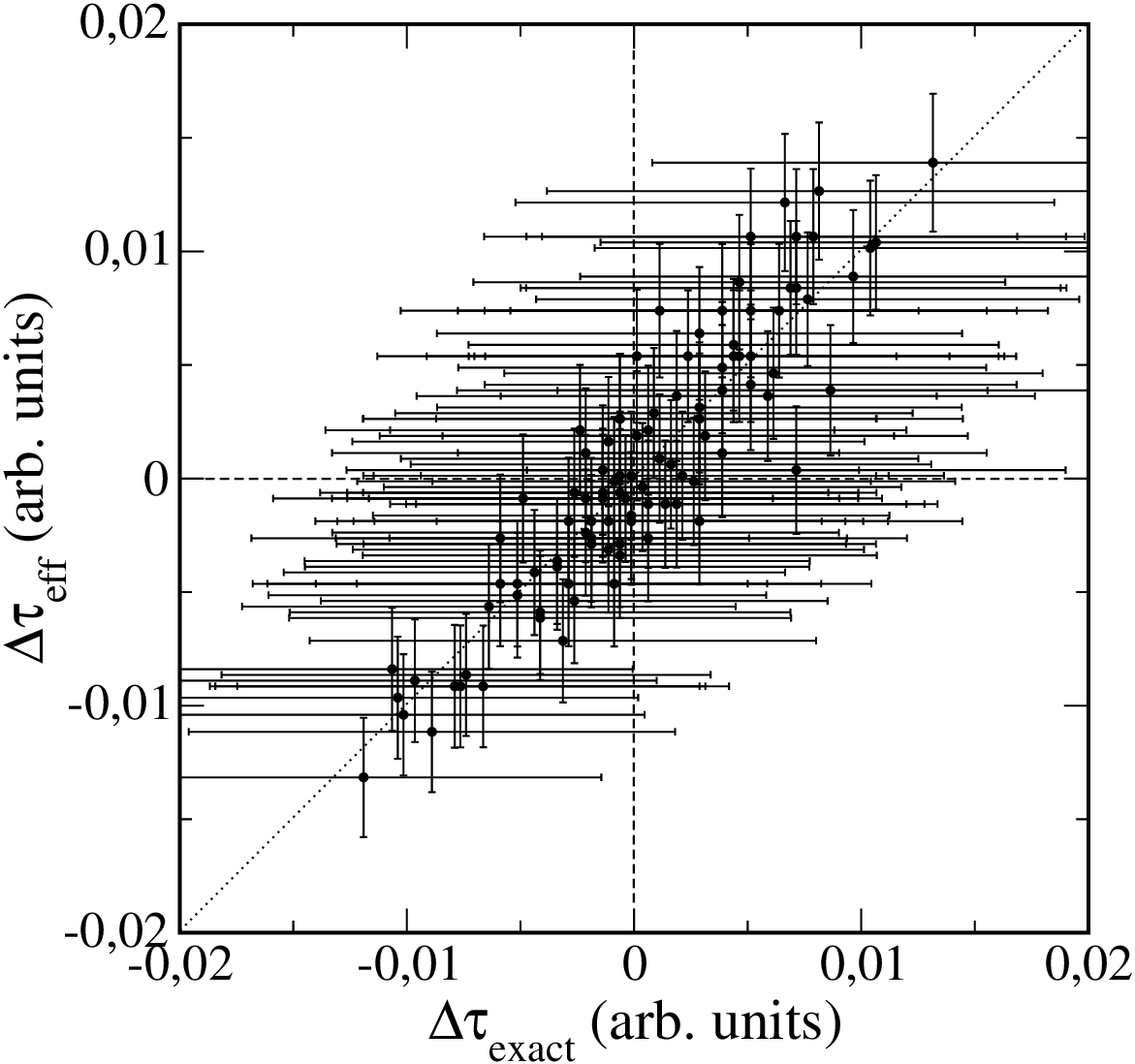}\end{center}

\caption{\label{fig_bias}Error estimates and 95\% confidence intervals for
the lifetime recovered simultaneously by the exact method $\Delta\tau_{exact}$
and by the effective variance method $\Delta\tau_{eff}$ for 100 randomly
noised signals and IRF's ($\sigma_{h}=\sigma_{s}=0.01$).}
\end{figure}

\textbf{Bias.} All methods perform similarly with regard to the bias
on lifetime estimation (Fig. \ref{fig_bias}). In this figure, we
reported the estimation by the ''effective variance'' method as
function of the estimation by the exact method. The biases of both
methods are highly correlated and practically identical. However,
underestimation of the confidence intervals by the approximate method
results in inconsistent estimations, i.e. it fails significantly more
than the exact method to include the exact value inside the confidence
interval, and confidence intervals for different realizations of the
noise are frequently disjoint. In this regard, Eq. \ref{result} performs
much better.

\begin{figure}[ht]
\begin{center}\includegraphics[%
  clip,
  scale=0.4]{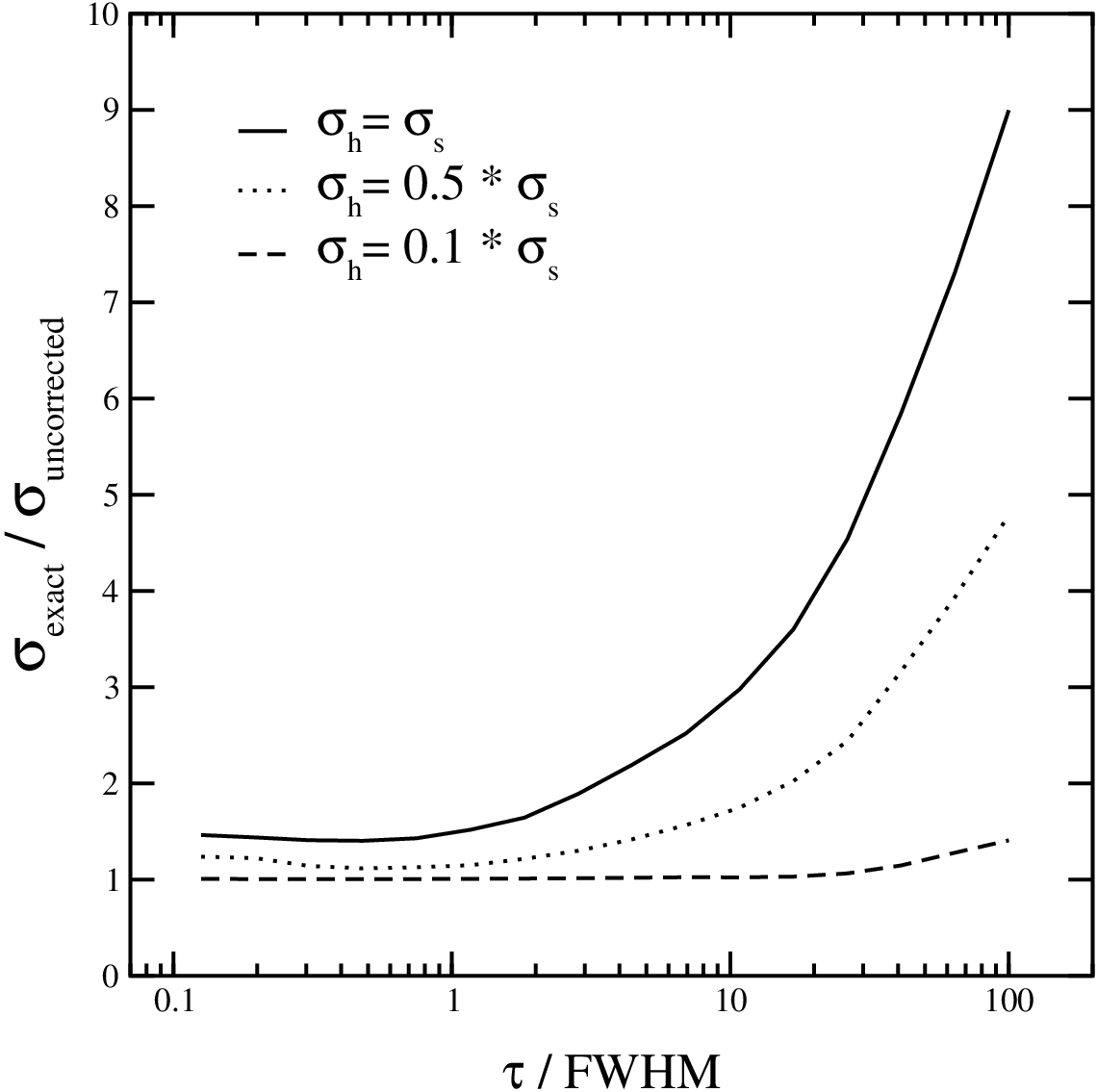}\end{center}

\caption{\label{cap:Ratio-sig}Ratio of the standard deviations of the posterior
pdf for the exact model ($\sigma_{exact}$) and for the uncorrected
model ($\sigma_{uncor}$) as a function of the theoretical lifetime.}
\end{figure}

\textbf{Accuracy of the IRF.} For a given lifetime, when the IRF is
measured with a better accuracy ($\sigma_{h}<\sigma_{s}$), the differences
observed between the various methods tend to vanish (Fig.\ref{cap:Ratio-sig}).
For instance, if the IRF is ten times more accurate than the signal,
the uncorrected method provides exact results over all the practical
range of lifetimes. It is also observed that longer lifetimes are
relatively more affected than shorter ones, which is a pure effect
of noise correlation (see next section).

\subsection{Structure of the correlation matrix}

The convolution of the mono-exponential model by the IRF is a vector
$\tilde{\mathbf{m}}$ which elements obbey the following reccurence\begin{equation}
\tilde{\mathbf{m}}_{i}=\textrm{exp}(-\frac{\Delta t}{\tau})\tilde{\mathbf{m}}_{i-1}+\mathbf{\hat{h}}_{i}+\mathbf{e}_{h,i}.\end{equation}
As soon as the IRF vanishes the correlation between consecutive points
is\begin{equation}
<\tilde{\mathbf{m}}_{i},\tilde{\mathbf{m}}_{i-1}>=\textrm{exp}(-\frac{\Delta t}{\tau})\end{equation}
As there is supposedly no correlation in the signal noise, the covariance
matrix $\mathbf{K}$ preserves this correlation scheme. An approximation
of the correlation matrix can thus been expressed as\begin{equation}
\mathbf{C}=\left(\begin{array}{ccccc}
1 & \rho & \rho^{2} & \cdots & \rho^{n}\\
\rho & 1 & \rho & \cdots & \rho^{n-1}\\
\vdots & \vdots & \ddots & \cdots & \vdots\\
\vdots & \vdots & \vdots & 1 & \rho\\
\rho^{n} & \rho^{n-1} & \cdots & \rho & 1\end{array}\right),\end{equation}
where $\rho=\textrm{exp}(-\frac{\Delta t}{\tau})$. The noise correlation
decays thus exponentially with de delay between points of the model,
the decay rate being the inverse of the theoretical lifetime.

\subsection{Support length of the IRF}

\begin{figure}[ht]
\begin{center}\includegraphics[%
  clip,
  scale=0.4]{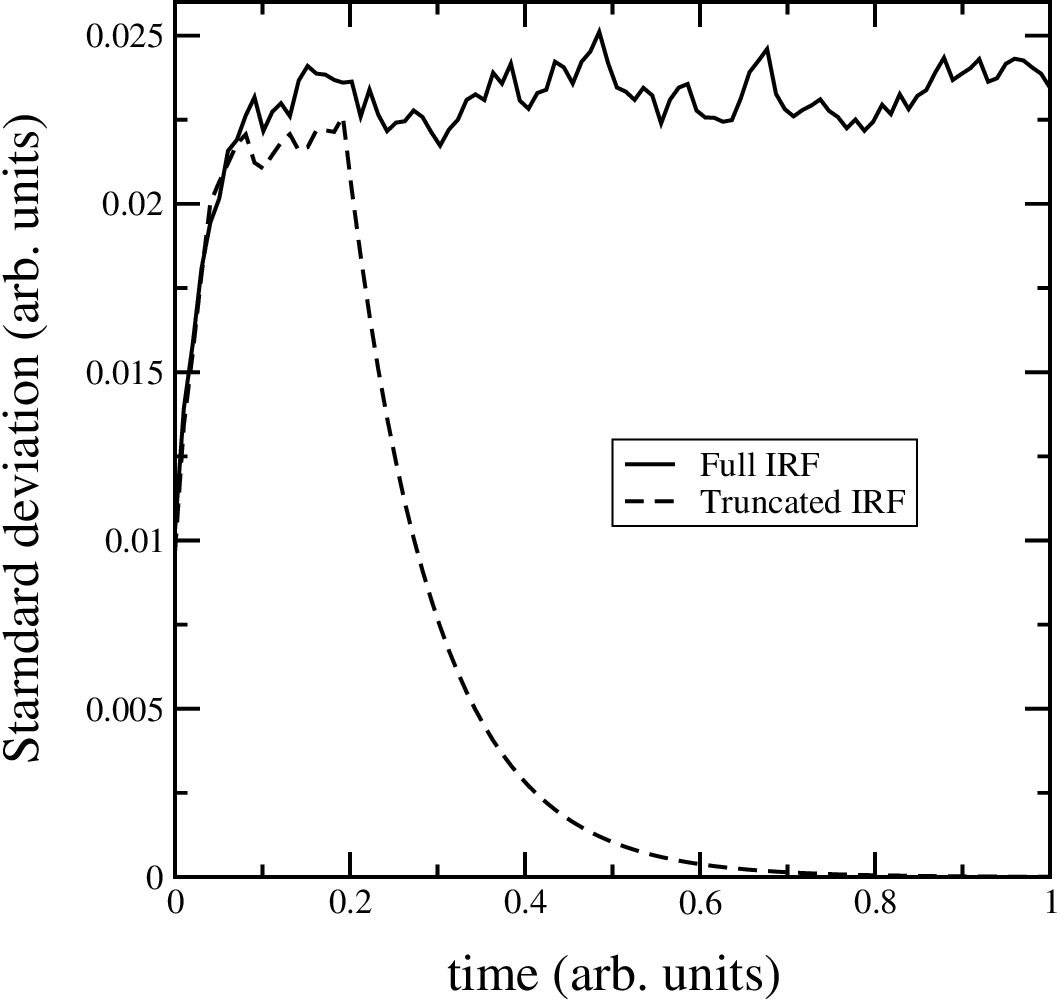}\end{center}

\caption{\label{cap:Standard-deviation}Standard deviation of the model, after
convolution with the IRF. Full line : IRF with full support; dashed
line : IRF with support limited to $t\leq0.2$. Same parameters as
for Fig. \ref{fig_dist1}.}
\end{figure}

When the IRF is recorded on the same support as the signal, most of
the its elements are pure noise. We saw above that these points contribute
significantly to the correlation of the noise in the convolved model.
Limiting the support of the IRF ($n_{h}<n$), or zeroing it's purely
noisy elements might thus enable to improve the correlation matrix.
If we observe the standard deviation for the convolved model (Fig.
\ref{cap:Standard-deviation}), we see that the truncation of the
support of the IRF contributes significantly to reduce the uncertainty
at larger times. As shown on Fig.\ref{fig_dist1}(c), this enables
some uncertainty reduction for the lifetime estimation, but the effect
of the correlated noise is still quite marked.

\subsubsection{Identifiability}

\begin{figure}[ht]
\begin{center}\includegraphics[
  clip,
  scale=0.4]{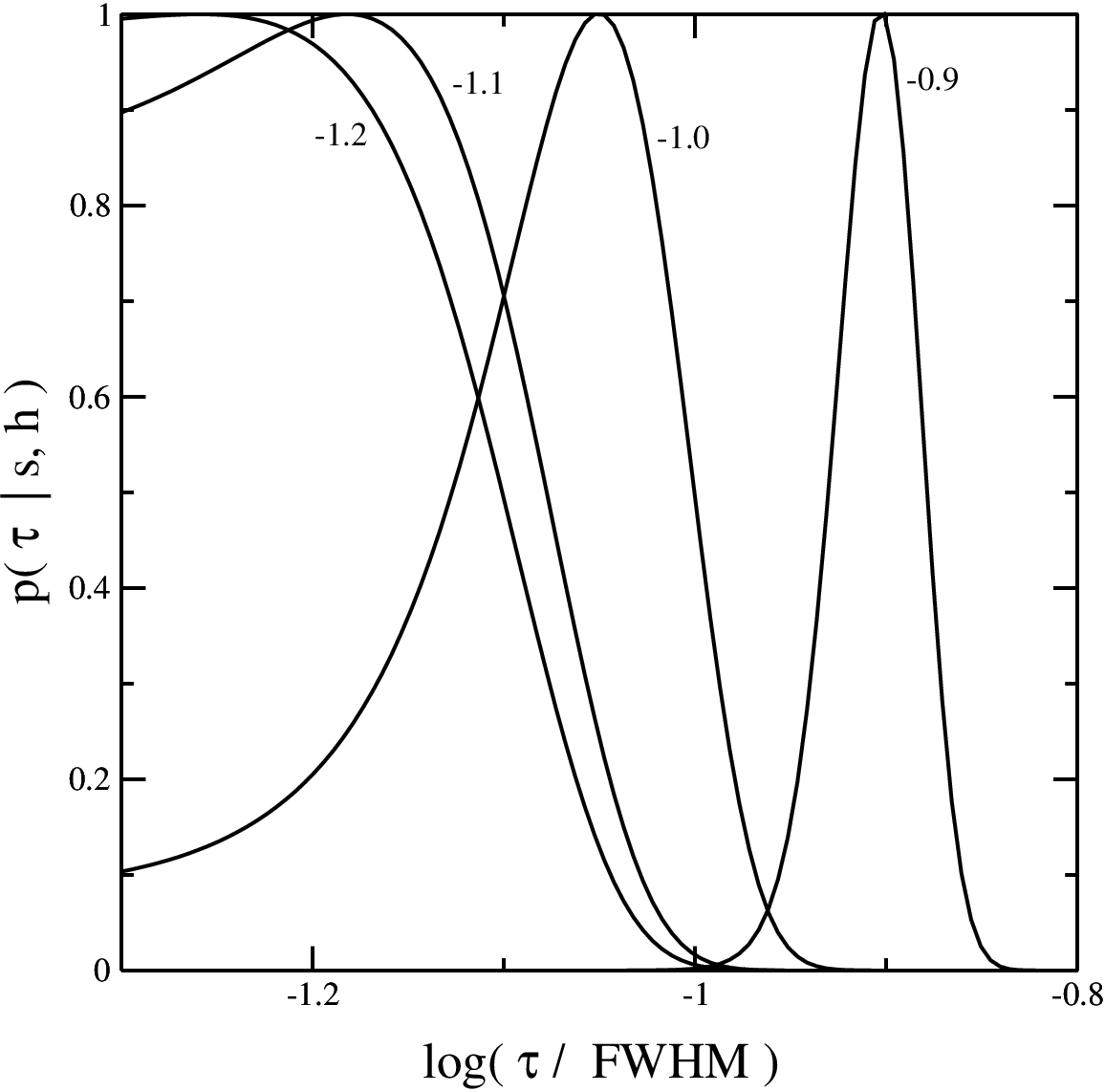}\end{center}

\caption{\label{cap:Identif}Evolution of the (unnormalized) posterior pdf
for $\tau$ at the resolution limit of the experimental setup. The
exact values for $\tau$ are reported alongside the curves. The standard
deviation for signal and IRF is $\sigma_{s}=\sigma_{h}=0.005$.}
\end{figure}

The behaviour of the present model with regard to the limits of detection
of lifetimes due to the IRF has been tested by reconstructing the
posterior pdf from synthetic signals generated with very small lifetimes.
The posterior pdf displays explicitely the non identifiability of
lifetimes that are too small (Fig. \ref{cap:Identif}). When $\tau$
decreases, the pdf becomes asymmetric, defining an upper limit for
the lifetime, but no lower limit, except the one imposed by the prior. 

\section{Conclusion}

The use of measured instrumental response functions for data deconvolution
is a source of uncertainty. We derived a new expression of the likelihood
within a bayesian framework to explicitely incorporate this effect
and display it's importance. Convolution of a noisy IRF with a model
curve produces a noisy model curve with correlated noise. 

This has been illustrated on a luminescence lifetime measurement setup,
for which it was shown that existing approximate treatments were markedly
defficient. It was also shown that, in this case, the correlation
length of the noise was directly related to the lifetime to be estimated.
Longer lifetimes are thus counterintuitively more affected by IRF's
uncertainty that shorter ones. Although the most efficient way to
reduce this effect is clearly to improve the IRF's measurement accuracy,
we have shown that an qualitative improvement can very simply be obtained
by zeroing those parts of the IRF consisting of pure noise.

The method has been applied to an homoscedastic noise pattern, but
extension to cases where the noise is dependent on signal intensity
(e.g. photon counting methods) is straightforward, as long as the
normal noise distribution approximation is valid. Similarly, cases
where the IRF is locally fluctuating due to minor modifications of
the experimental setup can be easily treated by a careful modelling
of the variance/covariance matrix.

We are studying extension of this method to Poisson uncertainties, and to the
evaluation of the resolution limits of a fluorescence TCSCP apparatus
\citep{Livesey87}. The ultimate goal is to obtain consistent uncertainty estimation 
for lifetimes recovered from fluorescence spectra analysis.

An alternative treatment is to model the IRF by a function, which
parameters pdf's are estimated by a bayesian analysis\[
p(\mathbf{m}|s,\mathbf{R}_{s},\mathbf{h},\mathbf{R}_{h})=\frac{p(\mathbf{m})}{p(\mathbf{s})}\int d\mathbf{p}_{h}\: p(\mathbf{s}|\mathbf{m},\mathbf{R}_{s},\mathbf{p}_{h})p(\mathbf{p}_{h}|\mathbf{h},\mathbf{R}_{h}).\]

\bibliographystyle{unsrt}

\end{document}